# Ferroelectric superdomain controlled graphene plasmon for tunable mid-infrared photodetector with dual-band spectral selectivity

Junxiong Guo[1,*], Lin Lin[2], Shangdong Li[3], Jianbo Chen[1,4], Shicai Wang[2], Wanjing Wu[1], Ji Cai[1], Tingchuan Zhou[4], Yu Liu[5,*], Wen Huang[2,6]

1. School of Electronic Information and Electrical Engineering, Chengdu University, Chengdu 610106, China
2. School of Electronic Science and Engineering (National Exemplary School of Microelectronics), University of Electronic Science and Technology of China, Chengdu 610054, China
3. School of Electronics and Information Technology (School of Microelectronics), Sun Yat-sen University, Guangzhou 510006, China
4. College of Material Science and Engineering, Sichuan University, Chengdu 610064, China
5. School of Integrated Circuits, Tsinghua University, Beijing 100084, China
6. Yangtze Delta Region Institute (Huzhou), University of Electronic Science and Technology of China, Huzhou 313001, China

* Corresponding authors. E-mail: guojunxiong@cdu.edu.cn (J.X. Guo) and y-liu-17@mails.tsinghua.edu.cn (Y. Liu)

# Abstract

Dual-band infrared photodetectors (DBIPs) can discriminate desired signals from complex scenes and thus are highly expected for threat-warning, remote sensing, and astronomy applications. Conventional DBIPs with high performances are, however, typically based on semiconductor thin films, but remain the challenges of expensive growth and cooling requirements. Here, we report a room-temperature graphene plasmonic photodetector with tunable dual-band infrared spectral selectivity driven by ferroelectric superdomain. The periodic ferroelectric polarization array with nanoscale ring shapes provides an ultrahigh electrostatic field for spatially doping of monolayer graphene to desired patterns, and is further used to excite and confine intrinsic graphene plasmons. Our devices exhibit tunable resonance photoresponse in both two bands of 3.7~16.3 μm and 15.1~52.1 μm. The numerical calculations show that our devices own ultrahigh responsivities of 667~1080 A $W^{-1}$ at room temperature in the range of 5~50 μm. Our devices make possible the applications of the infrared imaging system and both stationary and motion states of objects detection. These investigations provide a novel approach for advanced infrared system construction by employing a simple, low-cost, uncooled multispectral detectors array.

# 1. Introduction

Multispectral systems [1-4] have capabilities to distinguish both absolute temperature and unique signatures from complex scenes, also enable enhancing sensitivity via signal processing algorithms, making them highly desirable for advanced infrared detection. Applications, including threat-warning, remote sensing, and astronomy have been extended by employing the technique of multi-bands infrared focal plane arrays [5, 6]. In particular, dual-band photodetectors [7, 8] may screen out spurious information, such as background interference in the infrared region. Many commercial devices for multiband infrared detection have been presented [9, 10], which can operate by using sequential and simultaneous mode (such as HgCdTe quantum well detectors) or imaging techniques. Due to the high sensitivity and complementary metal–oxide–semiconductor (CMOS) technical maturity, these devices are widely used for mid-infrared (MIR, 3.0~50 μm, ISO 20473 standard) lights detection. However, the high-performance commercial photodetectors mainly rely on expensive growth thin-films and inevitably face the limitation of cooling requirements. The goal to develop a novel photodetector with low-cost fabrication and uncooled work temperate is not only the scientific pursuit, but also the industrial requirement.

Recently, the emerging two-dimensional materials have been explored for optoelectronic applications including display electrodes, laser, optical modulator, photovoltaic and plasmonic devices [11-14]. Among these materials, tremendous efforts have been concentrated to develop graphene based plasmonic photodetectors because of their distinct characteristics [15, 16]. First, the gapless graphene enables charge carriers producing over an ultrawide spectrum, ranging from ultraviolet to terahertz frequencies, for light detection [17]. Furthermore, graphene possesses the features that the carrier density can be easily reconfigured via electrostatic doping, and the tunable surface plasmon can be excited with much lower energies and confined with higher efficiencies than those of noble metals [18]. Free of complex spatial alignment and expensive growth technique, nor requiring cooling equipment for multispectral systems, a variety of prototype infrared photodetectors based on graphene plasmonics with tunable spectral photoresponse, operating at room temperature, have already been demonstrated in a single device [14, 15, 19, 20]. High-performance plasmonic infrared detectors are, however, usually based on pattered graphene, but they inevitably suffer from the edge-disorder and further degrade the device mobility [16]. Although the alternative approach of integrating metal plasmonic nanostructures with graphene has been developed, it only enables a tunable selective light filtering and detection within visible and near-infrared ranges (0.3~3 μm) by varying the nanostructured geometry or the applied gate voltage [14, 21]. Additionally, previous graphene plasmonic photodetectors have mainly focused on single-band tunability and face the common challenges of high operating voltage and complex fabrication, which are not suitable for the practical industrial requirement in complex senses.

In this article, we use a ferroelectric superdomain, alternating up- and down-polarized ferroelectric domains with nanoscale-wide ring shape, to excite and tune graphene plasmons for room-temperature infrared detection. By reversibly writing the geometries of superdomain and varying the graphene chemical potential, our designed devices feature tunable spectral response in both blue (3.7~16.3 μm) and red (15.1~52.1 μm) bands. We also propose an infrared imaging application using integration monolayer graphene on ferroelectric domain. Our devices are capable to operate at zero-bias of source/drain voltage without external gate electrodes, and exhibit an ultrahigh photoresponsivity up to 1080 A W$^{-1}$ at room temperature.

## 2. Tuning of graphene plasmon by ferroelectric polarization

Graphene plasmon, a collective form of electron excitation, allows to be utilized for tunable plasmonic devices. Efficiently excitation and tuning of graphene plasmons, relying on the carrier concentration with lights, however, are still a key challenge. The ferroelectric polarization can offer an equivalent electric field of $10^7$~$10^{10}$ V m$^{-1}$ [22], which is several orders higher than that of traditional gate-produced electrostatic field and making them extremely attractive candidates for doping of graphene to obtain the desired carrier concentration. To estimate the effective carrier concentration of graphene, it is necessary to convert the ferroelectric-polarization-induced doping of graphene into a conventional back-gating measurement. Similar to the usually applied gate voltage, the equivalent voltage ($V_{equ}$) of ferroelectric remanent polarization ($P_r$) creates an electrostatic potential difference ($\phi$) between graphene and ferroelectric layer, and the increased charge carriers result in a shift of chemical potential ($\mu_c$). Therefore, $V_{equ}$ can be described by

$$V_{equ} = \frac{\mu_c}{e} + \phi \tag{1}$$

where $\mu_c/e$ and $\phi$ are determined by the graphene chemical (quantum) capacitance and geometric capacitance ($C_g$), respectively [23]. The Fermi level ($E_F$), namely chemical potential, in graphene shifts as a function of carrier density ($n$), following $\mu_c = E_F = \hbar|v_F|\sqrt{\pi n}$, where $|v_F|$ is the Fermi velocity of 1.1 ×10$^6$ m s$^{-1}$ [23]. For ferroelectric-gated graphene device, $\phi = ne/C_{FG}$ and $V_{equ} = \eta P_r/C_{FG}$, where $C_{FG}$ and

$\eta$ denote geometric capacitance of ferroelectric gate and overall effective factor of ferroelectric-induced equivalent voltage applied on graphene, respectively. Combined with equation (1), we can get

$$\frac{\eta P_r}{C_{FG}} = \frac{\mu_c}{e} + \frac{e\mu_c^2}{C_{FG}\pi\hbar^2 v_F^2} \qquad (2)$$

In general, the remanent polarization ($P_r$) of ferroelectrics ranges from 10 to 180 µC cm$^{-2}$ [22]. From the pioneer works [24, 25], we set $\eta = 0.05$ cm$^{-2}$ V$^{-1}$, and $C_{FG}$ ranges from 2 to 5 µF cm$^{-2}$. By numerical calculating, the graphene chemical potential shifts in a range of 74~790 meV (Fig. 1a), depending on ferroelectric polarization and geometric capacitance.

We note the optical properties are strongly dominated by the chemical potential (also Fermi level). Therefore, the optical behaviors of graphene can be dynamically tuned by ferroelectric doping, following a transition from dielectric to metallic feature. Combined with the random-phase approximation (RPA), the dynamic conductivity of graphene ($\sigma_g(\omega)$) can be derived from the Kubo formula [26, 27], which is given by the interband and intraband contributions

$$\sigma_g(\omega) = \sigma_{intra} + \sigma_{inter,1} + i\sigma_{inter,2} \qquad (3)$$

where the intraband conductivity ($\sigma_{intra}$) follows Drude-like behavior

$$\sigma_{intra} = \sigma_0 \frac{4\mu_c}{\pi} \frac{1}{\hbar\tau_1 - i\hbar\omega} \qquad (4)$$

where $\sigma_0 = \pi e^2/(2h)$, $\mu_c > 0$, and $\tau_1$ is the relaxation rate for intraband transitions. The interband conductivity behaves

$$\sigma_{inter,1} = \sigma_0 \left(1 + \frac{1}{\pi}\arctan\frac{\hbar\omega - 2\mu_c}{\hbar\tau_2} - \frac{1}{\pi}\arctan\frac{\hbar\omega + 2\mu_c}{\hbar\tau_2}\right) \qquad (5)$$

$$\sigma_{inter,2} = -\sigma_0 \frac{1}{2\pi} \ln\frac{(2\mu_c + \hbar\omega)^2 + \hbar^2\tau_2^2}{(2\mu_c - \hbar\omega)^2 + \hbar^2\tau_2^2} \qquad (6)$$

where $\tau_2$ is the relaxation rate for interband transitions. In the longwave MIR to terahertz frequencies, the spectral response is mainly imparted to a Drude peak by following equation (4), because of the decisive contribution of intraband free carriers absorption [16]. Therefore, the relaxation rate for interband part is estimated as 0. The relaxation rate for intraband transitions can be achieved by $\tau_1 = \hbar/\tau$, and $\tau$ is the relaxation time following $\tau = \mu_e\mu_c/ev_F^2$, where $\mu_e$ is the electron mobility which is estimated as 9000 cm$^2$ V$^{-1}$ s$^{-1}$. For example, if the chemical potential of graphene induced by ferroelectric superdomain is 0.54 eV, we can obtain the relaxation time and relaxation rate for intraband transitions are 0.54 ps and 1.22 meV, respectively.

As a result, by employing the ferroelectric-doping induced graphene chemical potential shifts, we can attain a tunable dynamic conductivity with different frequencies. Specifically, the imaginary part of graphene conductivity ($\sigma_{g,i}$) can be easily tuned to positive and negative values (Fig. 1b), depending on chemical potential and incident wavelengths. When $\sigma_{g,i} > 0$, graphene follows an ultra-thin metal-like behavior and supports a transverse-magnetic (TM) surface plasmon polaritons (SPP) wave transportation, however, it stops supporting the TM SPP wave when $\sigma_{g,i} < 0$ [28].

For integrating monolayer graphene with ferroelectric film, it behaves p-doping graphene on down-polarized domain and near-intrinsic graphene on up-polarized domain [25]. In the MIR region, the imaginary part values for near-intrinsic and typical p-doped graphene are negative and positive from Fig. 1b, respectively. That is, free of the edge-disorder in patterned graphene and beyond the complex fabrication and dielectric breakdown of traditional electrostatic gating using local electrodes, the adjacent ferroelectric domains with opposite polarized direction (180°-phase-difference) offer a non-destructive approach to spatially modulating carrier behaviors in graphene to locally confine the TM SPP transportation. To achieve the expected patterning of ferroelectric polarization, some advanced switching technologies have been developed, such as direct writing with nanoscale resolution using piezoresponse force microscope (PFM) probe [22] and large-scale printing using water-induced ferroelectric switching [29].

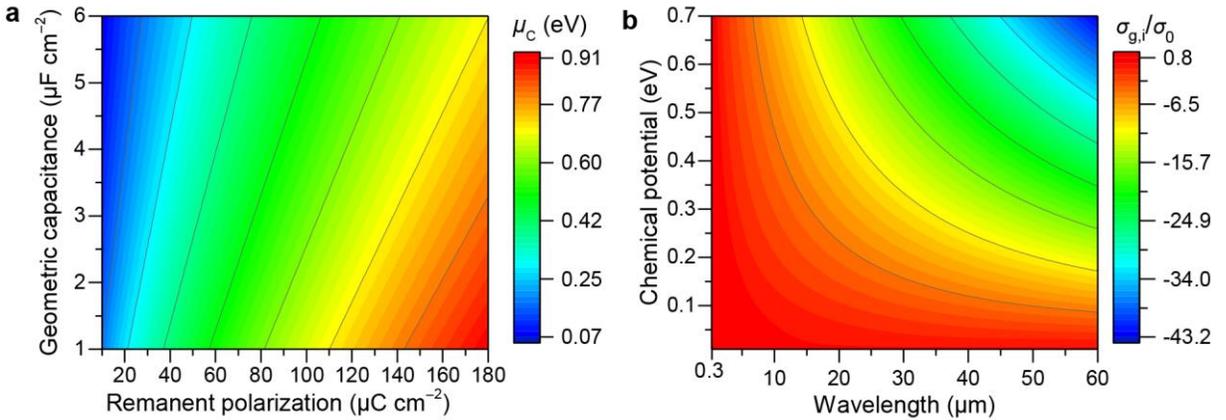

**Figure 1.** Doping of graphene by ferroelectric polarization.
(**a**) Ferroelectric polarization induced graphene chemical potential as a function of the geometric capacitance and remanent polarization. (**b**) Imaginary part of the reduced graphene conductivity ($\sigma_{g,i}/\sigma_0$) as a function of the chemical potential and frequency at room temperature. $\sigma_0$ is obtained by $\sigma_0 = e/4\hbar$.

# 3. Dual-band photodetector
## 3.1. Design and optical response of device

The core of our device consists of single-layer graphene (SLG) integrated on a periodically-polarized ferroelectric thin film, namely ferroelectric superdomain (Fig. 2a). The PFM probe with nanoscale tip is used for switching of ferroelectric polarization to the desired superpattern with adjacent upward and downward domains. To excite graphene plasmons in dual-band MIR frequencies, we first construct the ferroelectric polarization to ring-shaped downward domain array ($R_{in}$ and $R_{out}$ denote the inner and outer circles radius of ring, respectively), and other zones with different periods ($P$) are polarized to upward domains (Fig. 2a, b), resulting in doping of graphene charge carrier with p-doped and near-intrinsic behaviors (Fig. 2b).

The device simulations were performed using the finite element method (FEM). Within FEM models, graphene is described as surface current applied on ferroelectric, and the dynamic conductivity is derived from the Kubo formula using RPA method as above mentioned. In MIR frequencies, the intraband transition will be the only contributor to optical response when $|\mu_c| > \hbar\omega/2$, allowing for graphene plasmonic enhancement [16, 30]. The interband contribution is negligible and set zero value in calculations. The dielectric constants of ferroelectric thin film, including refractive index real and imaginary parts, are obtained from an online database [31], and an epitaxial growth ferroelectric of $BiFeO_3$ (25 nm thickness) with bottom electric layer of $La_{2/3}Sr_{1/3}MnO_3$ (25 nm thickness) on $SrTiO_3$ substrate (500 nm thickness) is employed in our device. The simulated electric-field intensity of the ring-shaped area ($E$) is much higher than that of other downward zones ($E_0$) (Fig. 2c). Interestingly, we note electric-field intensity in both the inner and outer circle borders are strongly enhanced compared to that of other zones within the ring-shaped downward domain. As a result, the transmission extinction ($1-T/T_0$) shows two resonance peaks in MIR region (Fig. 2d), where $T$ and $T_0$ denote the transmittances of graphene integrated on ferroelectric and bare ferroelectric thin film (inset of Fig. 2d), respectively. We fit the spectral data (hollow diamonds) using the Lorentz method, and the solid lines are the fitting results.

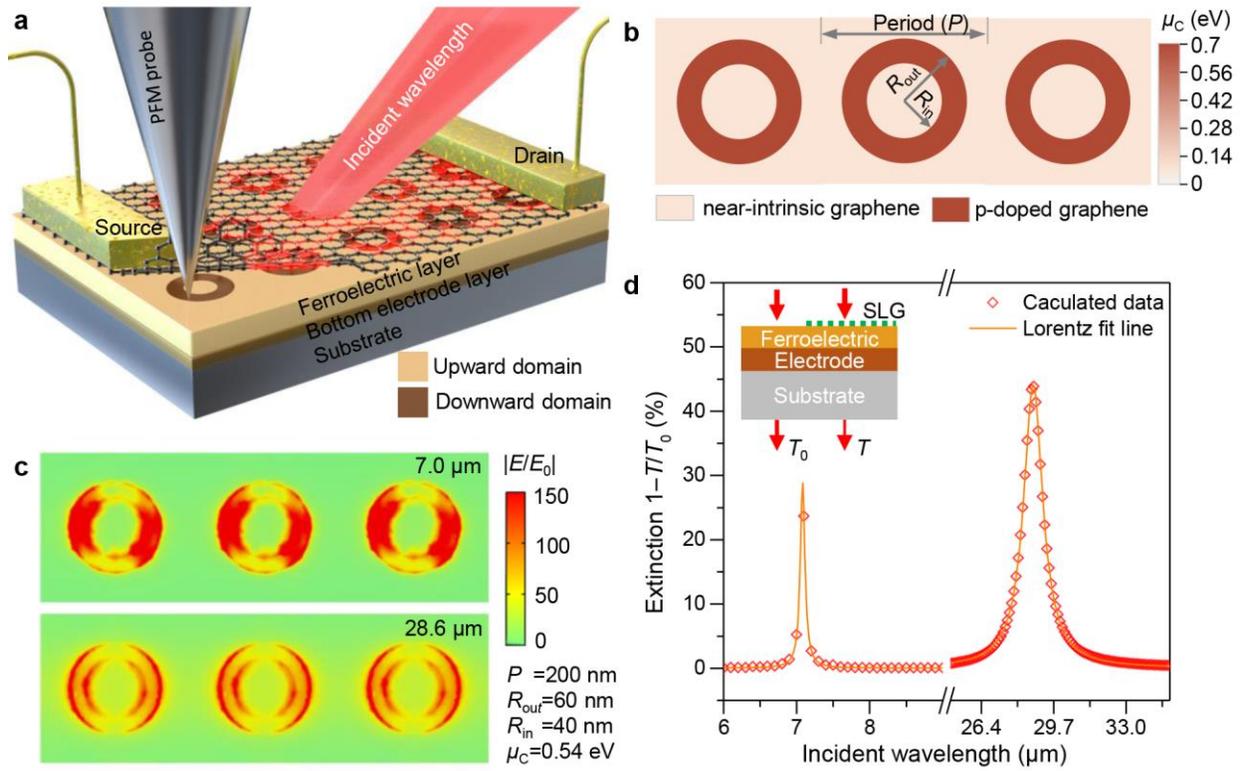

**Figure 2.** Dual-band photodetector using graphene plasmon driven by ferroelectric superdomain. (**a**) Illustration of device structure and operation principle. PFM probe is used to switch the ferroelectric thin film to desired superdomain structure. Two metal layers with different work function is deposited acting as source/drain electrodes. (**b**) Spatial doping of graphene by periodic ferroelectric polarization. The graphene on upward and downward domains follows near-intrinsic and typical p-doped behaviors, respectively. $R_{in}$, radius of inner circle with upward ferroelectric polarization; $R_{out}$, radius of outer circle with downward ferroelectric polarization. (**c**) Top-view of simulated electric field distribution at 7.0 and 28.6 μm wavelength illumination. The simulation is performed using a single unit and further make them to form a row. (**d**) Calculated results of transmission extinction $1-T/T_0$ (diamond squares) and corresponding fit results using Lorentz method (solid lines). $T$ and $T_0$ are the calculated transmittance of ferroelectric integrated graphene and bare ferroelectric, respectively.

## 3.2. Tunable photoresponse of device

The strong electric-field enhancement and transmission spectral signal resonance confirm the unprecedented potential of the device using ferroelectrically-controlled graphene plasmons to detect MIR light. For SPP in MIR region, the atomic thickness of graphene enables higher confinement and a lower dispersion loss compared with arrayed metal resonators [32, 33]. Fig. 3a, b presents the near-field distribution of local surface plasmon resonance (LSPR) models in ferroelectric-doping of graphene. The field hotspots are located along the borders of near-intrinsic and p-doped graphene on ferroelectric. By calculating the confinement percentage as a function of given distance $d$ (dashed line in Fig. 3a), $d_x$ and $d_y$ (dashed lines in Fig. 3c), we observe that 90% of mode energy ($z$-direction) is confined within 12 nm from graphene sheet (Fig. 3b) and 98% of mode energy ($xy$-plane) is confined in ring-shaped graphene on downward ferroelectric domain (Fig. 3d). In comparison, if we employ the gold array to confine 90% mode energy, the spread distance will be over 500 nm from the gold surface. These results confirm the stronger field confinement of graphene integrated on ferroelectric, resulting in a low dispersion loss and providing a novel approach for tuning of optical response within MIR frequencies.

We now turn to calculate the optical properties of the designed device. Fig. 3e plots the transmission extinction spectra, $1-T/T_0$, for ferroelectric-doping of graphene with chemical potential ranging from 0.1 to 0.7 eV. In all simulated devices, the period of polarized domain, radius of inner circle and outer circle of ring-shaped downward polarization are 200 nm, 20 nm and 30 nm, respectively. We observe four prominent features from Fig. 3e: 1) Two bands (band 1 and 2 mask with light-blue and light-red color, respectively) are significantly enhanced in each fixed chemical potential; 2) The resonance wavelength exhibits blue-shift with increasing graphene chemical potential, and the resonance peak position shifts with ferroelectric-induced graphene chemical potential are summarized in Fig. 3f; 3) The peak intensity increases apparently with chemical potentials; 4) The bandwidth narrows with increasing graphene chemical potential. In these devices, the dimension of alternately ferroelectric domains with opposite polarized directions (tens of nanometers) is much smaller than incident wavelengths, satisfying the demand of coupling the excited graphene plasmons with incident waves. By varying the ferroelectric-induced graphene chemical potential (0.1~0.7 eV), the resonance peak can be tuned in two regions of 6.1~14.4 μm (band 1) and 25.1~44.9 μm (band 2), respectively.

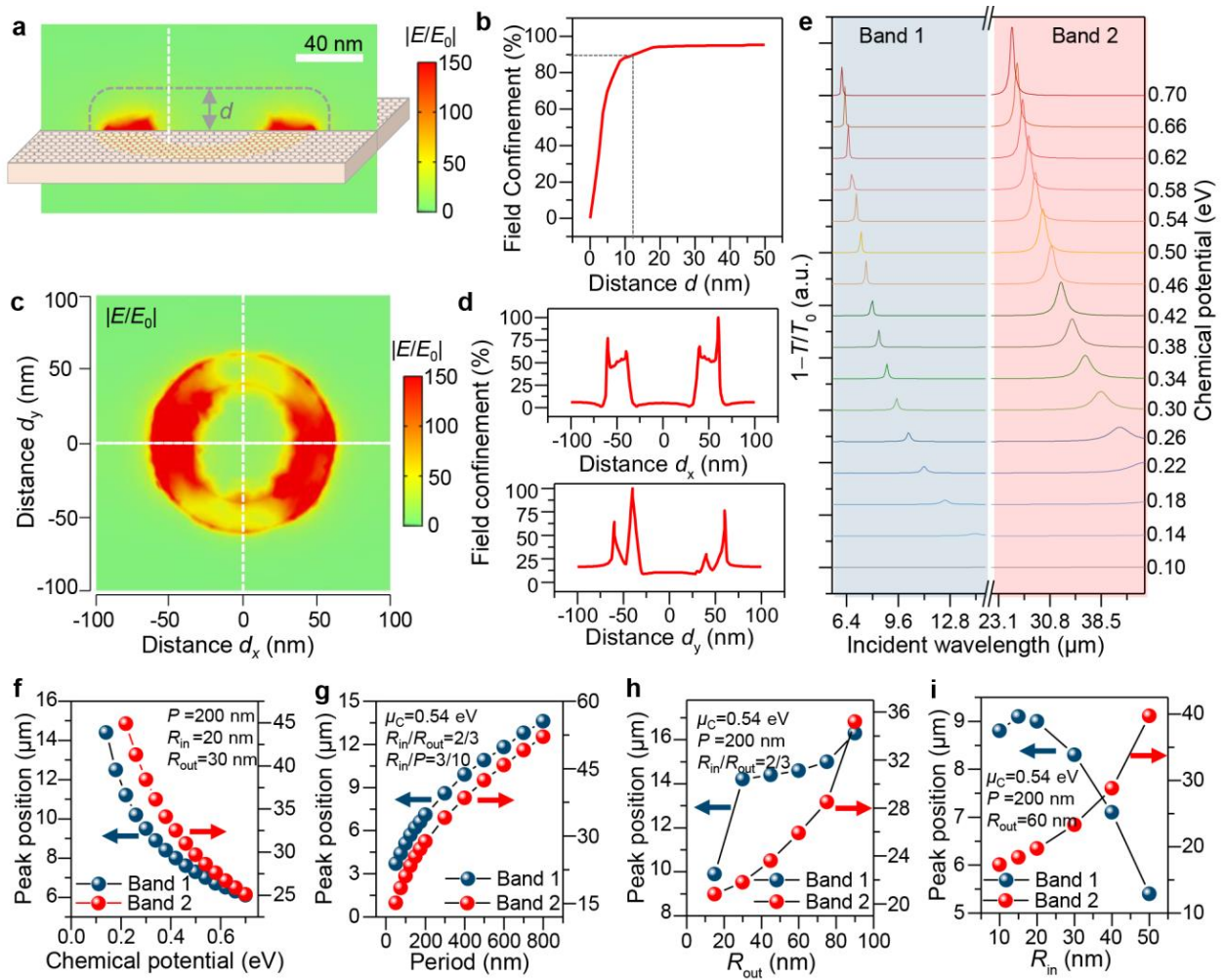

**Figure 3.** Tuning of graphene plasmons for dual-band photodetector.
(**a**) Simulation results (cross-section) of electric-field distribution under 7.0 μm wavelength illumination. $P$ = 200 nm, $R_{in}$ = 40 nm, $R_{out}$ = 60 nm and $\mu_c$ = 0.54 eV. (**b**) Percentage of electric-field intensity confined within a volume extending a distance $d$ outside the device, corresponding to the dashed line in **a**. (**c**) Top-view of simulated electric-field distribution as same as the device of **a**. (**d**) Similar to **b**, related electric-field confinement of graphene planes, corresponding to the dashed lines follow x- and y- directions in **c**, respectively. (**e**) Transmission extinction (1-$T/T_0$) spectra of device. Dual-band sensitive feature is observed and the resonance frequency can be tuned by varying the ferroelectric polarization induced graphene chemical potentials. The resonance peak positions of devices as functions of ferroelectric polarization induced graphene chemical potential of (**f**), polarized period of (**g**), outer circle radius of (**h**) and inner circle radius of (**i**), respectively.

We have also performed simulations for optimizing the geometric size of ferroelectric polarization to guide our device design. As a given chemical potential of 0.54 eV, we plot the resonance frequency position of devices as a function of different ring widths in Fig. 3g. By changing the period of designed devices from 50 to 800 nm, it enables a remarkable tunability of resonance peaks in both two ranges of 3.7~13.6 μm (band 1) and 15.1~52.1 μm (band 2). Similar results can be observed in Fig. 3h, the resonance peaks of the device varied ranging in two bands of 16.3~9.9 μm and 35.2~20.4 μm with scaling ring-shaped downward polarization (radius of the out ring) from 90 to 15 nm ($R_{in}/R_{out}$ = 2/3). If we fixed $\mu_c$ = 0.54 eV, $P$ = 200 nm, and $R_{out}$ = 60 nm, we found that a similar blue-shift occur in band 2, however, a red-shift occur in band 1 with scaling inner circle radius from 50 to 10 nm (Fig. 3i). Compared to the similar red-shifts in Fig. 3g, h, this abnormal blue-shift of resonance peak position in Fig. 3i may attribute to the relative width of ring-shaped down-polarized domain. In the cases of scaling the period and ring-shaped down-polarized domains, the relative width of ring (1−$R_{in}/R_{out}$) remains the same value of 1/3. When we vary the inner circle radius from 10 to 50 nm, the relative width of ring decreases from 5/6 to 1/6. With the thinner ring-shape of down-polarized domain, it matches a shorter incident wavelength to resonant the excited graphene plasmons. This result has also been demonstrated in previous works [20, 34], such as graphene nanoribbons and graphene disks. From these observations, we conclude that our devices can attain desired values of resonance peak position in MIR range by balancing the chemical potential of graphene and geometric size of polarized ferroelectric domain.

### 3.3. Photoelectric performance of detector

The responsivity of the photodetector ($R_{ph}$) is given as $R_{ph} = I_{ph}/P_{in}$, where $I_{ph}$ and $P_{in}$ are the photocurrent and incident power, respectively. For our device, the photocurrent can be derived from $I_{ph} = AV_{equ}\mu e \Delta n$, where $A$ is the effective area, $\mu$ is the charge carrier mobility and $\Delta n$ is the incident light produced effective carrier density. In MIR range, we can estimate the chemical potential shift ($\Delta \mu_c$) for graphene samples based on the Pauli blocking position according to the extinction spectra [34], shown in Fig 3e. For details to estimate the chemical potential shift, please refer to the electronic supplementary information. The actual charge carrier density can be derived from the chemical potential energy shift following $\Delta \mu_c = \hbar v_F \sqrt{\pi \Delta n}$ [23]. For higher incident frequencies (corresponding shorter wavelengths), there is a significant chemical potential upshift. It suggests that the photo-induced carrier density ranges from $1.4 \times 10^{13}$ to $3.1 \times 10^{13}$ cm$^{-2}$ (Fig. 4a).

To further realize the graphene plasmon for photocurrent enhancement, we also performed the photocurrent for a device consisting of graphene integrated on a single downward ferroelectric domain. From Fig. 4b, we observed the photocurrent density ($J_{ph}$), $J_{ph} = I_{ph}/A$, for the device of graphene on periodic polarized domains are sharply increased up to 667~1080 mA cm$^{-2}$, whereas the $J_{ph}$ for graphene on a single domain is only 2.2~5.4 mA cm$^{-2}$. Therefore, it suggests that the significantly enhanced $J_{ph}$ for the device of graphene on periodic domains is harnessed by the plasmon resonance, and the lower $J_{ph}$ for the device of graphene on single domain is similar to the uniform doping of graphene by conventional gating. From the above calculations, Fig. 4c plots the responsivity of our designed device dependence of incident wavelengths (5~50 μm), attaining ultra-high values in range from 667~1080 A W$^{-1}$. We could attribute this ultra-high responsivity to two main reasons. One is the electric field at interfaces enhanced by LSPR effect, which can accelerate the injection of hot carriers induced by incident light. Another is that the ferroelectric superdomain array is employed to control the graphene plasmons in our device. Benefiting from atomic smoothness of ferroelectric superdomain surface, graphene has shown significantly suppressed charge fluctuation and high carrier mobility. However, the defects of edge-disorder that could reduce the mobility of graphene and further limit the photocurrent in previous reports of plasmonic devices using patterned graphene.

The external quantum efficiency (EQE) is estimated by the number ratio of the collected charge carriers ($N_C$) to incident photons ($N_I$) and can be expressed as

$$\text{EQE} = \frac{N_C}{N_I} = \frac{hc}{e\lambda} R_{ph} \tag{7}$$

where $c$ is the vacuum light velocity and $\lambda$ is the incident light wavelength. As shown in Fig. 4d, the EQE of our device is ultra-high to 1652~26000% due to the amazing electric field enhancement induced by the strong LSPR effect.

We also perform the noise current ($i_N$) by following Johnson–Nyquist equipartition law. The noise current can be expressed as

$$i_N = \sqrt{\frac{4k_B T \Delta f}{R}} \tag{8}$$

where $k_B$ is the Boltzmann constant, $T$ is the temperature, $\Delta f$ is the bandwidth and $R$ is the resistance of graphene. For the giving bandwidth of −3dB in resonant peaks and room temperature of 300 K, we can obtain the noise current spectra as a function of incident wavelengths. Further, we can estimate the noise equivalent power (NEP) by using NEP = $i_N/R_{ph}$ and the specific detectivity ($D^*$) by using $D^* = \sqrt{A}/\text{NEP}$.

As shown in Fig. 4e,f, our designed device shows a low noise current with pA Hz$^{-1/2}$ scale and NEP with $10^{-12}$ to $10^{-14}$ W Hz$^{-1/2}$ scales in a wide range of 5~50 um. Moreover, the specific detectivity of our photodetector can achieve the highest value of up to $5.93\times10^{12}$ Jones (1 Jones = 1 cm Hz$^{1/2}$ W$^{-1}$) under the illumination of 5 μm at room-temperature.

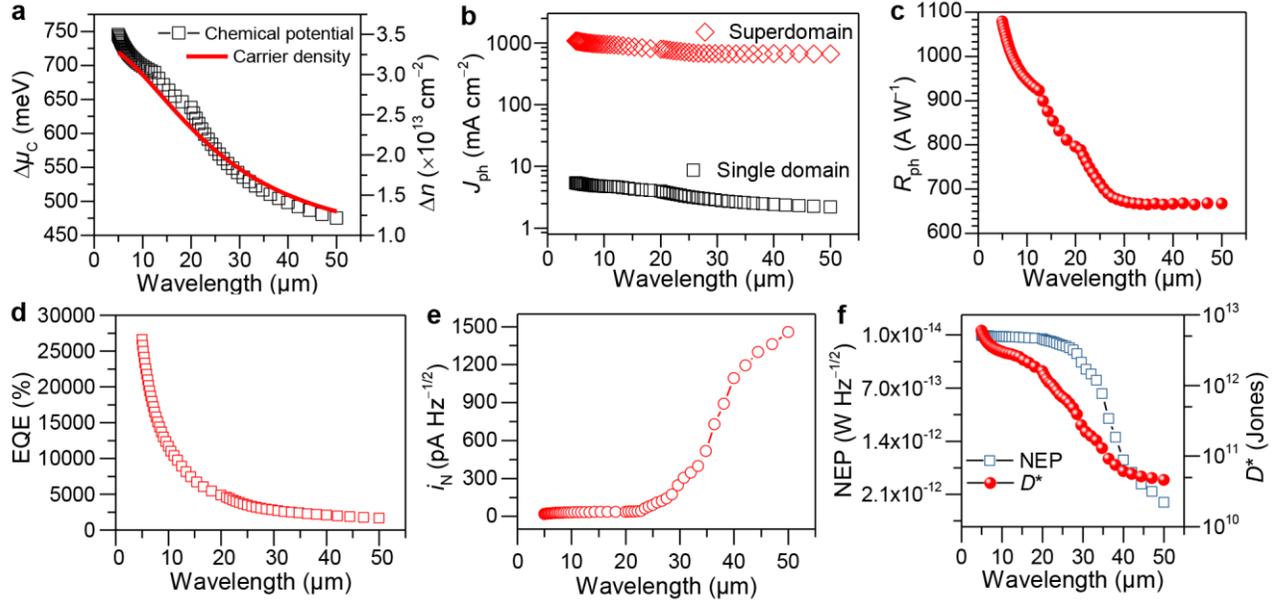

**Figure 4.** Photoelectric performance of dual-band photodetector. (**a**) Calculated graphene chemical potential shift ($\Delta\mu_c$) as a function of resonant wavelength (black hollow squares combined with a solid line) and related carrier-density excitation (red solid line). (**b**) Comparison of fabricated devices based on single-layer graphene integrated on ferroelectric superdomain (ring-shaped array, $P$ = 200 nm, $R_{in}$ = 40 nm, $R_{out}$ = 60 nm and $\mu_c$ = 0.54 eV, red squares) and single downward polarized ferroelectric domain (gray squares). (**c**) Calculated photoresponsivity ($R_{ph}$) as a function of resonant wavelengths. The incident power density is 1 mW cm$^{-2}$. (**d-f**) External quantum efficiency (EQE), noise current ($i_N$) spectra, noise equivalent power (NEP) and specific detectivity ($D^*$) as functions of incident wavelengths.

These results demonstrate the ability of the periodically polarized ferroelectric domain to excite and control graphene plasmon for providing stronger light detection. Compared to the current commercial multi-spectral devices, the performances of our photodetector are not the highest, but it can be operated at room-temperature, which is also one of the industrial pursuits. Besides, our designed devices have a simple fabrication route to integrate two-dimensional materials, including graphene, with thin-film substrate without traditional CMOS fabrication because of the van der Waals interactions. On the other hand, benefiting from the dual-band spectral selectivity, it can further improve the spectral sensitivity and tunability of photodetectors beyond state-of-the-art plasmonic detectors using patterned graphene and complex metallic array. These features demonstrated here, potentially offer a novel route for graphene plasmonic devices fabrication and provide a rich diversity for optoelectronic family.

### 3.4. Infrared imaging and warning applications

We finally performed the photodetectors to convert MIR light to electric signals and expect to integrate them into a system for practical applications. First, we can use the integrated photodetector array to image objects. For example, we can employ the proposed infrared system to show the image our hands temperature, the wavelength of human body radiation usually ranging from 4.5 to 13.5 μm with peaked around 7.5 μm [35], as illustrated in Fig. 5a. The operation is simply described as follows: First, the ferroelectric-integrated graphene plasmonic array is used to detect the infrared radiation produced by a hand with different wavelengths. In each unit of a single detector, both the geometric shape and remanent polarization intensity are given. The converted photocurrents are simultaneously gathered within the data collector. After processing the electric signals, a pseudocolor image will be displayed on a monitor. In this sense, each detector unit acts as pixel location and only requires a single-band operation, which is like the previous plasmonic using patterned graphene.

Our device is also can solve both the stationary and motional objects, illustrated the concept in Fig. 5b. Due to its feature of dual-band sensitivity for infrared radiation, our device can detect the motion states of human movements (such as walking in and out state) and the environment, simultaneously. Of course, our device can retain measurement signals even if the object does not move (environment, standing state), which only requires working at single-band mode. On the contrary, these motional scenes lose signal when the objects keep a stationary state, which is detected by traditional infrared sensors [36, 37]. By making full exploit of the features, our device also contributes its energy-saving property because of its nonvolatile doping of graphene and zero-bias operating voltage. Moreover, due to the marked difference of output signals (top panels in Fig. 5b), such as the photocurrent signal difference with incident wavelengths shown in Fig. 4b, our devices can be used as warning systems in the future by further signal processing.

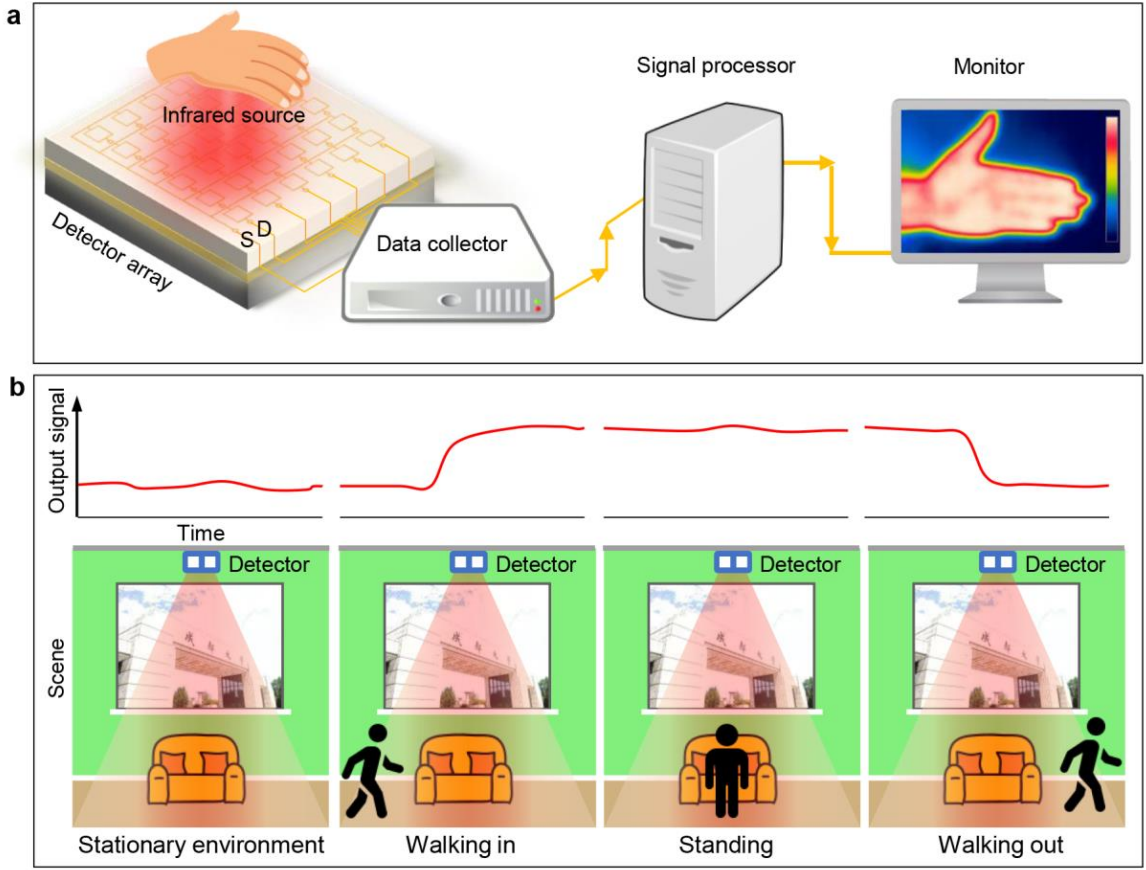

**Figure 5.** Applications of dual-band photodetector.
(**a**) Schematic illustration of integrated dual-band photodetector array on ferroelectric substrate for infrared imaging system. S/D, source/drain. Each detector unit works using single-band mode. (**b**) Ability to detect both stationary and motion state of humans in a single device. The detection of stationary (such as environment, standing state) and motional (such as walking in/out state) scenes operates using single- and dual-band, respectively.

## 4. Conclusions

We demonstrate a ferroelectric superdomain controlled graphene plasmons for tunable infrared detection in MIR frequency range operating at room-temperature. The remanent polarization with reversible switching feature provides a non-destructive method to construct desired doping patterns for graphene plasmons excitation and confinement. The electromagnetic simulations indicate that the excited graphene plasmon wave mainly propagates in the downward polarized domain, and resonance wavelengths can be easily tuned by varying ferroelectric-polarization-induced graphene chemical potential and scaling the geometric shape of periodically polarized domain. Interestingly, we have observed the designed device consisting of single-layer graphene integrated on periodically polarized ferroelectric with ring-shaped downward domain shows a tunable resonance photoresponse in both two bands of MIR lights. From numerical calculations, our device exhibits an ultrahigh responsivity of 1080 A W$^{-1}$ and specific detectivity of $5.93 \times 10^{12}$ Jones under the illumination of 5 μm. We also proposed an integrated ferroelectric-driven graphene photodetector array that could be used for infrared imaging and warning applications. These merits demonstrated here, intrinsic graphene plasmonics is controlled within periodic ferroelectric domain array, allowing for widely tunable plasmon resonance and light detection in the mid-infrared range, for a variety of crucial security, imaging applications.

## Acknowledgments

This work was financially supported by the National Natural Science Foundation of China (No. 61971108). J. Guo and Y. Liu thank Prof. J. Zhang and Dr. Y. Zhang from Beijing Normal University for fruitful discussions on ferroelectric superdomain construction.

## Conflict of Interest

The authors declare no conflict of interest.

## References

[1] J.M. Sun, M.M. Han, Y. Gu, Z.X. Yang, H.B. Zeng, Recent advances in group III-V nanowire infrared detectors, Adv. Opt. Mater. 6(18) (2018) 1800256.
[2] A. Rogalski, Recent progress in infrared detector technologies, Infrared Phys. Technol. 54(3) (2011) 136-154.
[3] F. Khodayar, S. Sojasi, X. Maldague, Infrared thermography and NDT: 2050 horizon, Quant. Infr. Therm. J. 13(2) (2016) 210-231.
[4] H.T. Lin, Z.Q. Luo, T. Gu, L.C. Kimerling, K. Wada, A. Agarwal, J.J. Hu, Mid-infrared integrated photonics on silicon: a perspective, Nanophotonics 7(2) (2018) 393-420.
[5] C. Downs, T.E. Vandervelde, Progress in Infrared Photodetectors Since 2000, Sensors 13(4) (2013) 5054-5098.
[6] A. Rogalski, J. Antoszewski, L. Faraone, Third-generation infrared photodetector arrays, J. Appl. Phys. 105(9) (2009) 091101.
[7] P. Wang, S.S. Liu, W.J. Luo, H.H. Fang, F. Gong, N. Guo, Z.G. Chen, J. Zou, Y. Huang, X.H. Zhou, J.L. Wang, X.S. Chen, W. Lu, F.X. Xiu, W.D. Hu, Arrayed van der Waals broadband detectors for dual-band detection, Adv. Mater. 29(16) (2017) 1604439.
[8] X. Tang, M.M. Ackerman, M.L. Chen, P. Guyot-Sionnest, Dual-band infrared imaging using stacked colloidal quantum dot photodiodes, Nat. Photonics 13(4) (2019) 277.
[9] A. Rogalski, Progress in focal plane array technologies, Prog. Quant. Electron. 36(2-3) (2012) 342-473.
[10] W. Lei, J. Antoszewski, L. Faraone, Progress, challenges, and opportunities for HgCdTe infrared materials and detectors, Appl. Phys. Rev. 2(4) (2015) 041303.
[11] C. Xie, C. Mak, X. Tao, F. Yan, Photodetectors based on two-dimensional layered materials beyond graphene, Adv. Funct. Mater. 27 (2017) 1603886.

[12] J.S. Ponraj, Z.Q. Xu, S.C. Dhanabalan, H.R. Mu, Y.S. Wang, J. Yuan, P.F. Li, S. Thakur, M. Ashrafi, K. McCoubrey, Y.P. Zhang, S.J. Li, H. Zhang, Q.L. Bao, Photonics and optoelectronics of two-dimensional materials beyond graphene, Nanotechnology 27(46) (2016) 462001.
[13] Y. Xie, B. Zhang, S. Wang, D. Wang, A. Wang, Z. Wang, H. Yu, H. Zhang, Y. Chen, M. Zhao, B. Huang, L. Mei, J. Wang, Ultrabroadband $MoS_2$ photodetector with spectral response from 445 to 2717 nm, Adv. Mater. 29(17) (2017) 1605972.
[14] F.H.L. Koppens, T. Mueller, P. Avouris, A.C. Ferrari, M.S. Vitiello, M. Polini, Photodetectors based on graphene, other two-dimensional materials and hybrid systems, Nat. Nanotechnol. 9(10) (2014) 780-793.
[15] T.J. Echtermeyer, S. Milana, U. Sassi, A. Eiden, M. Wu, E. Lidorikis, A.C. Ferrari, Surface plasmon polariton graphene photodetectors, Nano Lett. 16(1) (2016) 8-20.
[16] T. Low, P. Avouris, Graphene plasmonics for terahertz to mid-infrared applications, ACS Nano 8(2) (2014) 1086-101.
[17] F. Xia, H. Wang, D. Xiao, M. Dubey, A. Ramasubramaniam, Two-dimensional material nanophotonics, Nat. Photonics 8(12) (2014) 899.
[18] A.N. Grigorenko, M. Polini, K.S. Novoselov, Graphene plasmonics, Nat. Photonics 6(11) (2012) 749-758.
[19] D. Wang, A.E.L. Allcca, T.F. Chung, A.V. Kildishev, Y.P. Chen, A. Boltasseva, V.M. Shalaev, Enhancing the graphene photocurrent using surface plasmons and a p-n junction, Light-Sci. Appl. 9(1) (2020) 126.
[20] Q. Guo, R. Yu, C. Li, S. Yuan, B. Deng, F.J. Garcia de Abajo, F. Xia, Efficient electrical detection of mid-infrared graphene plasmons at room temperature, Nat. Mater. 17(11) (2018) 986.
[21] A.C. Ferrari, F. Bonaccorso, V. Fal'Ko, K.S. Novoselov, S. Roche, P. Bøggild, S. Borini, F.H. Koppens, V. Palermo, N. Pugno, Science and technology roadmap for graphene, related two-dimensional crystals, and hybrid systems, Nanoscale 7(11) (2015) 4598-4810.
[22] L.W. Martin, A.M. Rappe, Thin-film ferroelectric materials and their applications, Nat. Rev. Mater. 2(2) (2017) 16087.
[23] A. Das, S. Pisana, B. Chakraborty, S. Piscanec, S.K. Saha, U.V. Waghmare, K.S. Novoselov, H.R. Krishnamurthy, A.K. Geim, A.C. Ferrari, A.K. Sood, Monitoring dopants by Raman scattering in an electrochemically top-gated graphene transistor, Nat. Nanotechnol. 3(4) (2008) 210-215.
[24] Y. Zheng, G.X. Ni, C.T. Toh, C.Y. Tan, K. Yao, B. Zyilmaz, Graphene field-effect transistors with ferroelectric gating, Phys. Rev. Lett. 105(16) (2010) 166602.
[25] C. Baeumer, D. Saldana-Greco, J.M.P. Martirez, A.M. Rappe, M. Shim, L.W. Martin, Ferroelectrically driven spatial carrier density modulation in graphene, Nat. Commun. 6 (2015) 6136.
[26] S. Das Sarma, S. Adam, E.H. Hwang, E. Rossi, Electronic transport in two-dimensional graphene, Rev. Mod. Phys. 83(2) (2011) 407-470.
[27] E. Hwang, S. Adam, S.D. Sarma, Carrier transport in two-dimensional graphene layers, Phys. Rev. Lett. 98(18) (2007) 186806.
[28] A. Vakil, N. Engheta, Transformation optics using graphene, Science 332(6035) (2011) 1291-1294.
[29] Y. Tian, L. Wei, Q. Zhang, H. Huang, Y. Zhang, H. Zhou, F. Ma, L. Gu, S. Meng, L.-Q. Chen, C.-W. Nan, J. Zhang, Water printing of ferroelectric polarization, Nat. Commun. 9 (2018) 3809.
[30] M. Jablan, H. Buljan, M. Soljacic, Plasmonics in graphene at infrared frequencies, Phys. Rev. B 80(24) (2009) 245435.
[31] M.N. Polyanskiy, Refractive index database. https://refractiveindex.info. Accessed 2021-10-08.
[32] J.N. Anker, W.P. Hall, O. Lyandres, N.C. Shah, J. Zhao, R.P. Van Duyne, Biosensing with plasmonic nanosensors, Nat. Mater. 7(6) (2008) 442-453.
[33] D. Rodrigo, O. Limaj, D. Janner, D. Etezadi, F. Javier Garcia de Abajo, V. Pruneri, H. Altug, Mid-infrared plasmonic biosensing with graphene, Science 349(6244) (2015) 165-168.
[34] H. Yan, X. Li, B. Chandra, G. Tulevski, Y. Wu, M. Freitag, W. Zhu, P. Avouris, F. Xia, Tunable infrared plasmonic devices using graphene/insulator stacks, Nat. Nanotechnol. 7(5) (2012) 330-334.
[35] X. Shen, G. Ding, J. Wei, L. Zhao, Y. Zhou, H. Deng, L. Lao, An infrared radiation study of the biophysical characteristics of traditional moxibustion, Complement. Ther. Med. 14(3) (2006) 213-9.
[36] J. Yun, S.S. Lee, Human movement detection and identification using pyroelectric infrared sensors, Sensors 14(5) (2014) 8057-8081.
[37] J. Yun, M.H. Song, Detecting direction of movement using pyroelectric infrared sensors, IEEE Sens. J. 14(5) (2014) 1482-1489.